\documentclass[aps,prl,showpacs,twocolumn]{revtex4}
\usepackage{graphicx}

\newcommand{\be}{\begin{equation}}
\newcommand{\ee}{\end{equation}}
\newcommand{\bea}{\begin{eqnarray}}
\newcommand{\eea}{\end{eqnarray}}
\newcommand{\bqa}{\begin{eqnarray}}
\newcommand{\eqa}{\end{eqnarray}}
\newcommand{\bqs}{\begin{eqnarray*}}
\newcommand{\eqs}{\end{eqnarray*}}
\newcommand{\beq}{\begin{equation}}
\newcommand{\eeq}{\end{equation}}

\begin{document}

\draft
\noindent{{\it submitted}, Physical Review Letters \\ 
\title{ Equilibrium theory for a particle pulled by a moving optical trap}
\author{R. Dean Astumian}
\email{astumian@maine.edu}
\affiliation{University of Maine, Orono, Maine, USA}

\date{\today}

 \begin{abstract}
The viscous drag on a colloidal particle pulled through solution by an optical trap is large enough that on experimentally relavant time scales the mechanical force exerted by the trap is equal and opposite the viscous drag force.  The rapid mechanical equilibritation allows the system to be modeled using equilibrium theory, where the effects of the energy dissipation ({\em thermodynamic} disequilibrium) show up only in the coordinate transformations that map the system from the laboratory frame of reference, relative to which the particle is moving, to a frame of reference in which the particle is, on average, stationary and on which the stochastic dynamics is governed by a canonical equilibrium distribution function.  The simple equations in the stationary frame can be analyzed using the Onsager-Machlup theory for stochastic systems and provide generalizations of equilibrium and near equilibrium concepts such as detailed balance and fluctuation-dissipation relations applicable to a wide range of systems including molecular motors, pumps, and other nano-scale machines.
\end{abstract}

\maketitle

It is often emphasized that motion of a small particle through water under the influence of an external field is a thermodynamically non-equilibrium process.  This is of course true since there is a net flow of energy from the source of the field to the bath in which the particle moves and in which the energy is dissipated.  Thus it may seem that the title of this paper is a {\em nonsequitor} - how can an equilibrium theory possibly describe a patently non-equiibrium process ?  

What is overlooked is that for colloidal particles  (radius $< 10 \mu$m) on time scales long compared to the thermal (velocity) relaxation time ($\tau_v = m/R \approx 10^{-6} $s) the viscous drag is so large that acceleration is negligible (the over-damped limit) and the viscous drag force is equal and opposite the mechanical force \cite{purcell,ast_ajp06}.  This rapid {\em mechanical} equilibration allows a very simple description of the dynamical behavior of single particles interacting with moving potentials.   A coordinate transformation is used to shift the description of the system from the laboratory frame of reference, relative to which the colloidal particle is moving, to one in which the particle is, on average, not moving.  The stationary probability density function for the position of the particle on this new coordinate is given by a canonical equilibrium distribution function from which all thermodynamic quantities can be calculated.     Generalization of traditional thermodynamic equilibrium and near equilibrium concepts such as detailed balance and fluctuation-dissipation relations valid for systems arbitrarily far from thermodynamic equilibrium, but close to mechanical equilibrium, are also obtained.

The specific system discussed in this paper - a colloidal particle pulled along with a harmonic well that is caused to translate at constant velocity - is a particularly simple example that has been studied experimentally \cite{evans_prl} and theoretically \cite{jar_lanl,vanZ_pre03}.  The general perspective developed is that in most cases a nanoscale object in solution under the influence of  an external force field is itself in equilibrium and undergoes equilibrium fluctuations in a suitable coordinate system. The bath of course is not in equilibrium with the source of the field.  The particle serves as a ``conduit"  for energy to flow from the source to the bath, and the energy flow is modulated by the equilibrium fluctuations of the particle.  This perspective  is expected to apply to a wide range of systems, including not only colloidal particles but also to molecular motors and other nano-scale machines \cite{ast_pt02}.

Consider a colloidal particle pulled in one-dimension through solution by a  harmonic potential exerted by an optical trap moving at constant velocity (Fig. 1) as in the experiment of Wang et al. \cite{evans_prl}.   The equation of motion is \cite{vanZ_pre03}
\be
m \ddot{x} + R \dot{x} = -s (x - v t) + \epsilon(t) \,\,,
\ee
\noindent where $m$ is the mass of the particle, $R$ is the coefficient of viscous drag, $s$ is the spring constant of the harmonic optical trap, $v$ is the velocity at which the trap translates in space, and $\epsilon(t)$ is Gaussian white noise with statistical properties $<\epsilon(t)> = 0$ and $<\epsilon(t_1) \epsilon(t_2)> = 2 R k_B T \delta(t_2-t_1)$ where $k_B$ is the Boltzmann constant and T is the Kelvin temperature.    We can remove the explicit time dependence with a change of coordinate to
\be
\alpha = x - v(t - \tau_p) \,\,,
\ee
where $\tau_p = R/s  \,$ is the positional relaxation time.  At $\alpha = 0$ the system is in mechanical equilibrium where the force due to the trap is equal and opposite the drag force $s (x- v t) = R v$ when the particle moves at the velocity $v$ of the trap. Substituting 
\be
x = \alpha + v( t - \tau_p)
\ee
into Eq. (1) we have 
\be
m \ddot{\alpha} + R \dot{\alpha} + s \alpha = \epsilon(t)\,\,,
\ee
the equation for a damped harmonic oscillator with noise.  Subsequent to the velocity relaxation time $\tau_v = m/R$ the acceleration is negligible (i.e., the ``overdamped" limit) and we have
\be
 R \dot{\alpha} + s \alpha = \epsilon(t) \,\,.
\ee
Eq. (4) is precisely the equation studied by Onsager and Machlup \cite{onsager}.  The conditional probability density that the particle moves to position $\alpha_{\rm j}$ at a time $t + \Delta t$ given that it starts at position $\alpha_{\rm i}$ at time $t$ is  \cite{onsager}
\be
P(\alpha_{\rm j}, t+\Delta t|\alpha_{\rm i}, t) = \frac{\exp{\left[- \frac{s [ \alpha_{\rm j} - e^{- \Delta t/\tau_p} \alpha_{\rm i}]^2 }{2 k_B T (1 - e^{-2 \Delta t/\tau})}\right]}}{\sqrt{\frac{2  \pi k_B T}{s} (1 - e^{- 2 \Delta t/\tau_p})}} \,\,.
\ee
Eq. 5 is rather complicated.  The ratio of the conditional probability densities is much simpler:
\be
\frac{P(\alpha_{\rm j}, t+\Delta t|\alpha_{\rm i}, t)}{P(\alpha_{\rm i}, t+\Delta t|\alpha_{\rm j}, t)} =\exp\left[{\frac{s \left(\alpha_{\rm i}^2 - \alpha_{\rm j}^2\right)}{k_B T}}\right] \,\,.
\ee
Time ($\Delta t$) has disappeared entirely!  Similar relations were shown for arbitrary potentials using the Onsager/Machlup approach \cite{bier_pre99}.  Eq. 6 reflects the fact that the system obeys a detailed balance condition on the transformed coordinate
\be
p_{\rm eq}(\alpha_{\rm i}) P(\alpha_{\rm j}, t+\Delta t|\alpha_{\rm i}, t) = p_{\rm eq}(\alpha_{\rm j})\,\, P(\alpha_{\rm i}, t+\Delta t|\alpha_{\rm j}, t) \,\,.
\ee
The optical potential energy is
\be
U(x,t) = \frac{s}{2} (x - v t)^2 = \frac{s}{2} (\alpha - v \tau_p)^2 
\ee
from which we see that the term in the exponent in Eq. (7) can be written $s (\alpha_{\rm j}^2 - \alpha_{\rm j}^2) = \Delta U(t) + 2 R v \Delta \alpha$ where $\Delta U(t) = U(x_{\rm j}(t)) - U(x_{\rm i}(t))$, and $\Delta \alpha = \alpha_{\rm j} - \alpha_{\rm i}$.  Since this term depends only on the potential energy at the two endpoints, and on the distance between them $\Delta \alpha (= \Delta x)$, Eq. (7) can be rewritten in the laboratory coordinate as
\bea
\frac{P(x_{\rm j} , t+\Delta t|x_{\rm i}, t)}{ P(x_{\rm i}, t+\Delta t|x_{\rm j}, t)}= \exp{ \left[\frac{-\Delta U(t) - 2 R v \Delta x}{k_B T}\right]} \,\,, \\
\nonumber 
\eea

\noindent where  $\Delta x = x_{\rm j} - x_{\rm i}$.  The quantity $R v |\Delta x|$ is the minimum work needed to reliably move a distance $\Delta x$ at constant velocity $v$ through the viscous medium \cite{der_prl99}.   Eq. (10) is a generalized detailed balance condition for systems in mechanical equilibrium but for which no thermodynamic equilibrium state exists.  For $v=0$ Eq. (10) reduces to the standard detailed balance condition derived from knowledge of the thermodynamic equilibrium to which the system relaxes in the long time limit.
 
Eqs. (6), (7), and (10) with time broken into increments $\Delta t$ that are long compared to $\tau_v$ but short relative to $\tau_p$ can form the basis for a general treatment of over-damped stochastic dynamics using the variational approach developed by Onsager and Machlup \cite{onsager} which can be extended to arbitrary potentials \cite{ast_ajp06,bier_pre99} with  time dependence \cite{tarlie_pnas98}.  

Here we focus on results that hold for times long compared to $\tau_p$ after which the distribution on the transformed coordinate obeys the canonical stationary equilibrium probability density function \cite{onsager}
\be
p_{\rm eq}(\alpha) =  \frac{\exp{(-\frac{s\, \alpha^2}{2 k_B T})}}{\sqrt{(2 \pi k_B T/s)}} \,\,,
\ee
with the mean square displacement from $\alpha = 0$ given by a fluctuation dissipation relation
\be
\left<\alpha^2\right> = \sqrt{s/(2 \pi k_B T)} \int_{-\infty}^{\infty} d\alpha \alpha^2 \exp{(-\frac{s \, \alpha^2}{k_B T})} = \frac{  k_B T}{ 2 s}
\ee
and with the two time auto-correlation function \cite{risken_fpeq}
\be
\left<\alpha(t_1)\alpha(t_2)\right> = \frac{ k_B T }{2 s} \exp{\left[-(t_1-t_2)/\tau_p\right]} \,\,.
\ee
The distribution function $p_{\rm eq}(\alpha)$ is Gaussian.  This is because the moving potential is  harmonic (i.e., quadratic).  There is a tendency toward misidentification of a Gaussian distribution with a thermodynamically near-equilibrium distribution and vice versa.  This is not correct.  For a strongly asymmetric potential such as the asymmetric sawtooth often used in discussions of Brownian motors \cite{ast_prl04,pro_prl04, mag_prl03} the distribution function is not well approximated by a Gaussian even at thermodynamic equilibrium.  On the other hand, with the moving harmonic trap used here, the distribution is Gaussian after $t > \tau_p$ irrespective of the velocity $v$ of the trap, i.e., irrespective of how far the system is from thermodynamic equilibrium.  The question of whether a distribution function for some quantity is or is not Gaussian has as much or more to do with the underlying symmetry of the system as it does with whether the system is or is not near thermodynamic equilibrium.

In the experiment of Wang et al. \cite{evans_prl} a  feedback loop was used to assure that the trap moved at constant velocity and data were collected only after the  average velocity of the particle had attained a stationary value (i.e., after a time $ t > \tau_p$).  The particle then rides along with the moving potential at the average velocity $v$ in the laboratory frame.  The average position in the tranformed frame is $<\alpha> = 0$ with an average power consumption
\be
{\cal P}_{\rm avg} = R v^2 \,\,.
\ee 
The experiment was carried out close to both velocity and positional equilibrium.  The trap constant used was fairly  small ($s \sim 10^{-7}$ N/m$ = 10^{-7}$ kg/s$^2$),  the radius of the particle was $r \sim  10^{-6} $m so the mass was $m \sim r^3 \rho_p \sim 10^{-15}$ kg (where $\rho \sim 10^3$ kg/m$^3$ is the density of the particle), and the viscous drag coefficient according to the Stokes formula was $R \sim 6 \pi \eta r \sim 10^{-8}$ kg/s, where $\eta \sim 10^{-3}$ kg/(m s) is the viscosity of water. The velocity at which the trap was translated was $v \sim 10^{-6}$ m/s.  The velocity relaxation time was $\tau_v = m/R \sim 10^{-7}$ s, the positional relaxation time was $\tau_p = R/s \sim 10^{-1} $s,  and the average power input was $\sim 2.5 k_B T/$s.  

 As the particle fluctuates about the mechanical equilibrium point $\alpha = 0$ the instantaneous power ${\cal P}_{\rm inst}$ needed to keep the trap moving at constant velocity fluctuates 
 \be
 {\cal P}_{\rm inst} = v \times s (x - v t)  = v \times s (\alpha - v \tau_p) \,\,,
 \ee
where $s (x - v t)$, the reactive force due to the particle acting on the potential,  is equal and opposite the force of the potential acting on the particle.  Thus the amount of  work supplied by the external source in a time $\Delta t$ is
\be
\Delta W = v s \int_t^{t+\Delta t}dt' (\alpha - v \tau_p)\,\,,
\ee
the mean value of which is 
\be
\left<\Delta W\right> = v^2 s \tau_p \Delta t = {\cal P}_{\rm avg} \Delta t \,\,.
\ee
There are fluctuations about the mean due to the fluctuations of the particle.  Because $\alpha$ is Gaussian distributed  $W$ is also Gaussian distributed.  Thus we need only to calculate the variance 
\bea
 &\left<\left(\Delta W - \left<\Delta W\right>\right)^2\right> = v s\left< \left(\int_t^{t+\Delta t} \alpha(t') dt' \right)^2\right> \nonumber\\ 
 \\
 &=  \int_t^{t+\Delta t}\int_t^{t+\Delta t}\left<\alpha(t_1)\alpha(t_2)\right> dt_1 dt_2 \nonumber
\eea  
to get the probability density function.  Using Eq. (13) for the correlation function under  the double integral in Eq. (18), and taking the leading term for large $t$, we find for the variance  
\be
 \left<\left(\Delta W - \left<\Delta W\right>\right)^2\right> = 2 k_B T {\cal P}_{\rm avg} \Delta t \,\,,
\ee
and the probability density for the work done by the source is
\be
P(\Delta W) = \frac{\exp{\frac{\left(\Delta W - {\cal P}_{\rm avg} \Delta t\right)^2}{2 k_B T {\cal P}_{\rm avg}\Delta t}}}{\sqrt{4 \pi  k_B T {\cal P}_{\rm avg} \Delta t}} \,\,.
\ee
The work probability density obeys a generalized fluctuation dissipation relation \cite{boch_physa81}
\be
P(-\Delta W)  = P(\Delta W) \exp{\frac{-\Delta W}{k_B T}}
\ee
and the concomitant equality
\be
\left<\exp{\frac{-\Delta W}{k_B T}}\right> = \int_{-\infty}^{\infty} \exp{\frac{\Delta W}{k_B T}} P(\Delta W) d \Delta W = 1 \,\,.
\ee
that follows from the fact that the work probability density is normalized $\int_{-\infty}^{\infty} P(-\Delta W) d \Delta W = 1$.  The fluctuation-dissipation form of Eq. (19) is obvious from the physical perspective - those trajectories in which the internal energy of the bath increases are relatively more likely than those trajectories in which the internal energy of the bath decreases \cite{boch_physa81}.   For over-damped systems the generalized fluctuation-dissipation relations follow almost immediately from Onsager and Machlups work \cite{onsager}.  Bochkov and Kuzovlev \cite{boch_physa81} used only microscopic reversibility in their derivation, which is consequently more general.  More recently, equations of the form of Eq. (19) have been termed Fluctuation relations \cite{evans_adv} and Eq. (20) a non-equilibrium work energy relation \cite{jar_prl97}.

From Eq. (19) it is apparent that it is more likely that work is done by the source on  the bath, than that the thermal fluctuations in the bath conspire to do work on the source.  However, for small $\Delta t$ there is an appreciable chance that the fluctuations actually reverse the net flow of energy and that the work done in the interval is negative, i.e., that the bath does work on the source, a so-called ``violation of the second law" \cite{evans_prl}.   We can calculate the probability for this to happen by integrating the work probability density function from $-\infty$ to $0$,
\be
P\left[\Delta W<0\right] = \int_{-\infty}^0 P(\Delta W) d \Delta W = \frac{1}{2} {\rm erfc}\left(\sqrt{\frac{{\cal P}_{\rm avg} \Delta t}{4 k_B T}}\right) \,\,.
\ee
A plot of $P(\Delta W<0)/\left[1 - P(\Delta W<0)\right]$ against $\Delta t$ is shown in Fig. 2.  The curve agrees well with the experimental result of Wang et al. \cite{evans_prl}.  It is not at all surprising that such reversals of energy flow occur, but it is perhaps counter-intuitive that they are evident for micrometer colloidal particles on time scales as long as seconds, many times the positional relaxation time of the particle.

By focussing on the mechanically equilibrated motion of small scale systems in water simple equilibrium relations can be obtained for the probability density functions in a coordinate system shifted from the laboratory frame of reference.  These relations contain all information concerning the thermodynamically non-equilibrium exchanges of energy between the external source, the system of interest, and the bath.  The resulting equations provide generalizations of relations derived from equilibrium thermodynamics such as detailed balance and fluctuation-dissipation relations.  The perspective offered here is expected to provide important insight into a wide variety of nanoscale systems including biologically important processes.   For proteins we expect for parameter values $s \sim 10^{-3} N/m$,  $r \sim 10^{-8} m$, $m \sim 10^{-21} kg$ and $R \sim 10^{-11} kg/s$ from which we find the velocity relaxation time $\tau_v  \sim 10^{-10} s$, and the positional relaxation time within a nanometer well, $\tau_p = R/s \sim 10^{-8} s$.    Consequently a theory based on mechanical equilibrium is very valuable for understanding the large scale motions involved in protein folding, nanometer stepping of biomolecular motors, and large conformational changes of signal- and other energy- transducing proteins, all processes that take place while the macromolecule is in mechanical equilibrium but is driven by thermodynamically far from equilibrium processes in the environment.  

The generalized fluctuation-dissipation theorems are complementary to work on nanoscale brownian motors \cite{ast_sci97,ast_pt02} and fluctuation driven transport \cite{rei_pr02}.   The generalized fluctuation-dissipation theorems focus on symmetries that are preserved even in the presence of external directed forces that drive the system out of thermodynamic equilibrium, while the ``brownian motor" mechanism shows how structural asymmetry can combine with directionless energy input and thermal noise to drive directed motion.  A brownian motor can be viewed as a molecule or nanoscale device that  serves as a conduit coupling two external sources to a heat bath in such a way that the flow of energy from the stronger source can rectify the occasional reversal of the flow of energy between the bath and the weaker source, allowing energy to be pumped from bath to do work on the weaker source.  The energy for the reversal  is provided by the stronger source,  but the mechanism takes advantage of the omnipresent fluctuations in the energy flows due to thermal or other sources of noise.   These fluctuation driven transport processes \cite{ast_pra89} share far more in common with the coupled transport processes described by Onsager \cite{onsagerII,onsager} than they do with the mechanisms of macroscopic motors and pumps.


\begin{thebibliography}{99}
\bibitem{ast_ajp06} R.D. Astumian, 
Am. Jour. Phys. {\bf 74}, 683-688 (2006).

\bibitem{purcell} 
E. Purcell,
3--11 (1977).

\bibitem{evans_prl} G. M. Wang, E. M. Sevick, E. Mittag, D. J. Searles, and
D. J. Evans, 
 Phys. Rev. Lett. {\bf 89}, 050601-1--4 (2002).
 
 \bibitem{jar_lanl} O. Mazonka and C. Jarzynski, e-print cond-mat/0210505 (1999).

\bibitem{vanZ_pre03} R. van Zon, E.G.D. Cohen, 
Phys. Rev. E {\bf 67}, 046102 (2003).

\bibitem{ast_pt02}
R. D. Astumian and P. Hanggi, 
Phys. Today {\bf 55} (11), 33--39 (2002).

\bibitem{onsager} L. Onsager and S. Machlup, 
Phys. Rev. {\bf 91}, 1505--1512 (1953). 

\bibitem{bier_pre99}
M. Bier, I., Derenyi, M. Kostur, and R.D. Astumian, Phys. Rev. E.,
 Phys. Rev. E. {\bf 59}, 6422-6432 (1999).
 
 \bibitem{der_prl99}
 I. Derenyi, M. Bier, and R.D. Astumian, Phys. Rev. Lett. {\bf 83}, 903 (1999).
 
\bibitem{tarlie_pnas98}
M. Tarlie and R. D. Astumian,
Proc. Natl. Acad. Sci. USA {\bf 95}, 2039--2043 (1998).

\bibitem{mag_prl03}
M.O. Magnasco, Phys. Rev. Lett. {\bf 71}, 1477-1480 (1993).

\bibitem{ast_prl04}
R.D. Astumian and M. Bier, Phys. Rev. Lett. {\bf 72}, 1766 (1994).

\bibitem{pro_prl04}
J. Prost, J.F. Chawin, L. Peliti, and A. Ajdari, Phys. Rev. Lett. {\bf 72}, 2652 (1994).

\bibitem{risken_fpeq}
H. Risken, ``The Fokker-Planck Equation", Springer-Verlag, Berlin 1989.

\bibitem{boch_physa81}
G. N. Bochkov and Yu. E. Kuzovlev, 
Physica A{\bf 106}, 443--479 (1981).

\bibitem{evans_adv}
D. Evans and D. Searles, 
Adv. Phys. {\bf 51}, 1529-1585 (2002).

\bibitem{jar_prl97}
C. Jarzynski, 
Phys. Rev. Lett. {\bf 78}, 2690--2693 (1997).

\bibitem{ast_sci97} R. D. Astumian, 
Science {\bf 276}, 917--922 (1997).

\bibitem{rei_pr02}
P. Reimann, Phys. Rep. {\bf 361}, 57  (2002).

\bibitem{ast_pra89}
R.D. Astumian, P.B. Chock, T.Y. Tsong, and H.V. Westerhoff, Phys. Rev. A  {\bf 39}, 6416 (1989).

\bibitem{onsagerII}
L. Onsager, Phys. Rev. {\bf 37}, 405 (1931).

\end{thebibliography}
\end{document}